\documentclass[english,superscriptaddress,twocolumn]{revtex4}
\usepackage{times}
\usepackage[T1]{fontenc}
\usepackage[latin1]{inputenc}
\usepackage{graphicx}

\makeatletter

\usepackage{babel}
\makeatother
\begin{document}


\title{Raman-modes of index-identified free-standing single-walled carbon nanotubes}

\author{Jannik C. Meyer}

\thanks{These authors contributed equally to this work}

\affiliation{Max Planck Institute for Solid State Research, Stuttgart, Germany}

\author{Matthieu Paillet}

\thanks{These authors contributed equally to this work}

\author{Thierry Michel}

\affiliation{Laboratoire des Verres, Colloïdes et Nanomat\'eriaux (UMR CNRS
5587), Universit\'e de Montpellier
II, France}

\author{Alain Mor\'eac}

\affiliation{Groupe Mati\`ere Condens\'ee et Mat\'eriaux (UMR CNRS
6626), Universit\'e de Rennes I, France}

\author{Anita Neumann}

\author{Georg S. Duesberg}

\affiliation{Infineon Technologies Corporate Research, Munich, Germany}

\author{Siegmar Roth}

\affiliation{Max Planck Institute for Solid State Research, Stuttgart, Germany}

\author{Jean-Louis Sauvajol}

\affiliation{Laboratoire des Verres, Colloïdes et Nanomat\'eriaux (UMR CNRS
5587), Universit\'e de Montpellier
II, France}

\begin{abstract}

Using electron diffraction on free-standing single-walled carbon nanotubes
we have determined the structural indices (n,m) of tubes in the diameter range
from 1.4 to 3~nm. On the same free-standing tubes we have recorded Raman spectra
of the tangential modes and the radial breathing mode. For the smaller diameters (1.4-1.7~nm) these measurements confirm previously
established radial breathing mode frequency versus diameter relations, and would be consistent with the theoretically predicted proportionality to the inverse diameter. However, for extending the relation to larger diameters, either a yet unexplained environmental constant has to be assumed, or the linear relation has to be abandoned.  

\end{abstract}
\maketitle
\hyphenation{nano-tube nano-tubes}

%
%

Raman spectroscopy is an important technique in the characterization of carbon
nanotubes \cite{Dresselhaus_PhysicalPropsOfCNTs1998,sreich_CNTs-concepts+prop2004}. The characteristic features of the Raman spectrum of single-walled carbon nanotubes
(SWNTs) depend on the nanotube structure, defined by the
indices (n,m) \cite{DresselhausPhysRep05}. The so-called radial breathing mode
(RBM) is a fingerprint of single-walled nanotubes, and its frequency is related to the nanotube diameter.
The relation of the RBM frequency  $\omega_{RBM}$ to the nanotube diameter d
is often given as
 $\omega_{RBM}$=A/d+B. This relation well agrees with various calculations and
experiments \cite{Dresselhaus_PhysicalPropsOfCNTs1998,sreich_CNTs-concepts+prop2004,KuertiPRB98,SanchezPRB99,DresselhausPhysRep05,FantiniPRL04b,TelgPRL04,JorioPRL01,JorioPRB05},
however, the actual values found for A and B vary significantly.
The value of B is interpreted as an effect of the environment (substrate, bundle, or detergent) and is expected to be zero for free-standing nanotubes.

Up to now,
there exists no independent determination of the nanotube structure and diameter
in combination with its Raman spectrum or the RBM frequency on the same individual nanotubes. We present Raman spectroscopy
in combination with an independent determination of the nanotube structure by
electron diffraction \cite{GaoSWNTnanoareaED03,IijimaHelicity96,AmelinckxED98,LambinHelixCalc96,CowleyED97,ColomerSWNTUniqueHel04}. From the RBM frequencies measured on precisely identified nanotube structures, we obtain an RBM vs. diameter relationship that does not depend on any modelization of nanotube electronic or mechanical properties.

The experiments are based on a new simple procedure to create arbitrary nanostructures
by electron beam lithography in such a way that access by TEM is possible.
The structures, with the carbon nanotubes embedded, are created on
the edge of a cleaved substrate and made free-standing with an etching
process.

\begin{figure}
\includegraphics[%
  width=8.5cm]{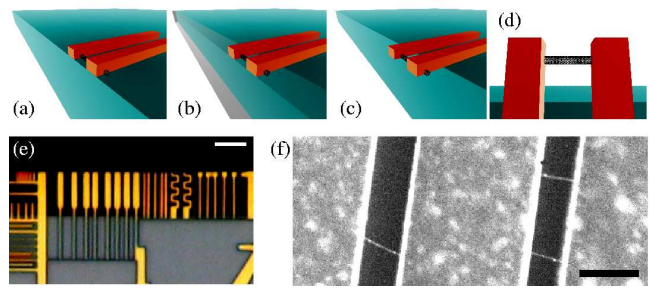}
\caption{(color online) Sample preparation procedure: (a) The substrate is
cleaved through a metallic grid which is on top of the carbon nanotubes.
(b) The sample is etched so that the structure is mainly undercut
from the side, removing the shaded volume. The resulting structure
(c) reaches out across the side edge of the substrate. Since the nanotube
is still held by the metal contacts, and the substrate is no longer
in the way, it is accessible for TEM investigations (d), viewed from
top. (e) is an optical microscope image of a free-standing structure (scale
bar 5\ensuremath{µ}m). The location of carbon nanotubes within the structure is known
from overview TEM images (f), so that micro-Raman measurements are
possible on precisely located nanotubes. (f) is a dark-field mode TEM image.
Scale bar is 1\ensuremath{µ}m. \label{cap:Schematic-illustration}}
\end{figure}

\begin{figure}
\includegraphics[%
  width=0.7\linewidth]{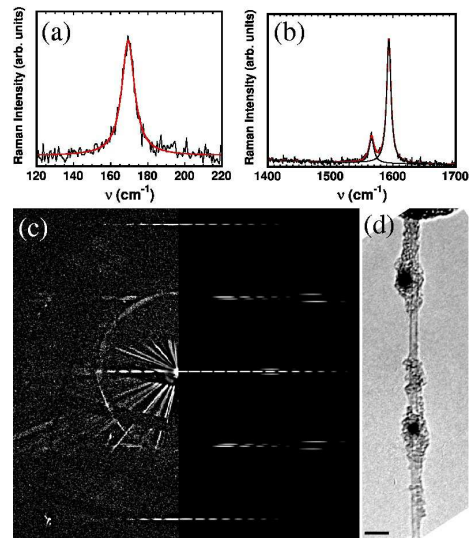}
\caption{(a) RBM and (b) TM ranges of the (11,10) SWNT (E$_{Laser}$=2.41~eV). (c) Diffraction pattern of this same
(11,10) nanotube (left: experimental, right: simulated image).
(d) High-resolution TEM image of the
same nanotube. Two small silver particles are visible, and at the
upper end is one of the contacts. Note that the most of the amorphous
carbon is deposited during the TEM analysis.\label{cap:2}}
\end{figure}

Single-walled carbon nanotubes are grown by chemical vapour deposition
(CVD) on highly doped silicon substrates with a 200~nm Silicon dioxide
layer \cite{PailletCVDJPCB04}. A metal structure consisting of 3~nm
Cr and 110~nm Au is created by electron beam lithography on top of
the as-grown carbon nanotubes. The substrate is then cleaved through
the metal grid structure. An etching process, as illustrated in Fig.
\ref{cap:Schematic-illustration}, is used to obtain freestanding
nanotubes: The sample is etched in a 30\% KOH bath at 60°C for 7 hours.
This removes quickly the bulk Si substrate, and slowly the oxide layer.
The etch rate of the doped silicon substrate can be controlled by
biasing it with respect to the bath. Since the oxide layer initially
acts as a mask, the structure is undercut mainly from the side of
the cleaved edge. An undercut of 10~$µ$m can be achieved, and the
etching process has to be stopped just when the oxide layer is completely
removed. After the etching process, the sample is transferred into
deionized water, isopropanol, and acetone before a critical point
drying step. On half of the free-standing nanotube samples, a tiny amount
of silver is deposited by thermal evaporation. The amount corresponding
to a 1~nm thick layer forms separated silver particles with a few nm
in diameter along the nanotube \cite{ZhangMetalDep00,ZhangMetalDep00b}.
Two small particles are visible in Fig. \ref{cap:2}d. We find
that the silver deposition on the nanotubes can lead to an
increase of the Raman intensity.

Since the substrate is no longer in the way, TEM is possible on the
free-standing part of the structure on the edge of the substrate.
The carbon nanotubes are held in place by the metal structure. Before
the micro-Raman experiments, overview TEM images are obtained at low
dose and voltage (60~kV) to get the position and orientation of
the carbon nanotubes with respect to the metal structure (Fig. \ref{cap:Schematic-illustration}f).

Since the metal structure is visible in the optical microscope and
the overview images show the nanotube location and their orientation with respect to the metal grid,
it is possible to carry out micro-Raman experiments on a perfectly localized and oriented
single tube. A first series of room-temperature Raman spectra were measured using the
Ar/Kr laser lines at 488~nm (2.54~eV), 514.5~nm (2.41~eV) and
647.1~nm (1.92~eV) in the back-scattering geometry on a triple substractive
Jobin-Yvon T64000 spectrometer equipped with a liquid nitrogen cooled
charge coupled device (CCD) detector. Another series of spectra were collected using a
tunable laser (1.57 eV-1.7 eV) with a Dilor XY800 spectrometer. In all the experiments
the instrumental resolution was 2~cm$^{-1}$. A precise positioning of the tubes under the laser spot (1 $\mu$m laser spot) was monitored with a piezoelectric nano-positioner. In our experimental
configuration, the incident light polarization is along the SWNT axis
(the Z axis), and no analysis of the polarization of the scattered
light is done.

After measuring the Raman spectra, diffraction patterns and high-resolution
TEM images of the same nanotubes investigated by Raman spectroscopy are obtained.
We record diffraction patterns on image plates in a Zeiss 912~$\Omega$ microscope operated at 60~kV.
The very straight and well-separated nanotubes obtained by our sample
preparation method allow a reliable analysis by electron diffraction.
The experimentally obtained diffraction pattern is compared with simulated
diffraction patterns, and
the nanotube indices and the incidence angle of the simulation are
adjusted until the simulated pattern matches the experimental one.
We verify that only exactly one pair of indices (n,m) matches the
experimental pattern, by checking that the simulated patterns for
all nearby indices clearly deviate from the experimental one for any incidence
angle. Details of our diffraction analysis procedure are given in \cite{MeyerSWNTED05}.
With an electron diffraction pattern that matches only a single pair of indices (n,m), the diameter is precisely known given the length of the carbon-carbon bond. Our diameter values are based on a C-C distance of 1.42 \AA.

After the Raman measurements, we have analyzed by electron diffraction
all the nanotubes that showed a Raman signal. Among these tubes,
three times the (11,10) nanotube was found. On the whole, we have obtained
spectra from 10 perfectly and unambiguously identified nanotubes: (11,10) (3 times), (15,6),
(16,7), (12,12), (17,9), (15,14), (27,4) and (23,21) (see additional materials).
Furthermore, we measured the spectra from two other tubes that could not be fully identified,
but of which diameters were quite precisely determined from the equatorial lines of their diffraction patterns,
i.e. 1.64 $\pm$ 0.05 nm and 2.3 $\pm$ 0.05 nm.

Figure \ref{cap:2}  shows the Raman spectra (using a 2.41~eV excitation) and the diffraction pattern of a (11,10) SWNT.
The exact determination of the transition energy of a SWNT requires the measurement
of the resonance profiles with a broad set of laser lines \cite{TelgPRL04,FantiniPRL04b}. However, the measurement of a detectable signal for the 2.41 eV incident energy means that the transition energy of the (11,10) SWNT is close to 2.41 eV.
As expected for an individual SWNT, the Raman spectrum is featured by a single narrow RBM located at 169.5~cm$^{-1}$ (FWHM = 7~cm$^{-1}$). The profile of the (11,10) TM bunch, displayed in Fig. \ref{cap:2}b,
is well fitted by using two Lorentzian components located at
1593.5~cm$^{-1}$ (FWHM = 6~cm$^{-1}$), and 1566~cm$^{-1}$ (FWHM = 7~cm$^{-1}$).
In our scattering geometry, the A symmetry modes are expected to contribute predominantly
to the Raman signal. The assignement of the two detected peaks on the (11,10) Raman spectrum as A modes
is coherent with \textit{ab initio} calculations \cite{DubaiPRL02} and
previously reported TM frequencies vs diameter plot \cite{JorioPRL03}.
Concerning the accuracy of our procedure of the localization of the tube,
it can be pointed out that we found three times the same RBM and TM spectra on three different isolated nanotubes identified as (11,10) from their electron diffraction patterns.
The reproduction of the Raman spectra validates our procedure of localization of the tubes.
It was recently shown that an uniaxial strain (which could occur on some of our
suspended nanotubes) can lead to a downshift of the TM frequencies \cite{CroninPRL04}.
From our TM frequencies we conclude that uniaxial strain (if present) is much
smaller in our samples than in \cite{CroninPRL04}. Nevertheless,
this region of the spectra is still under careful examination. In the following, we are only focusing on the RBM frequencies for which a small uniaxial strain has undetectable effects \cite{CroninPRL04}.

\begin{figure}
\includegraphics[%
 width=8.5cm]{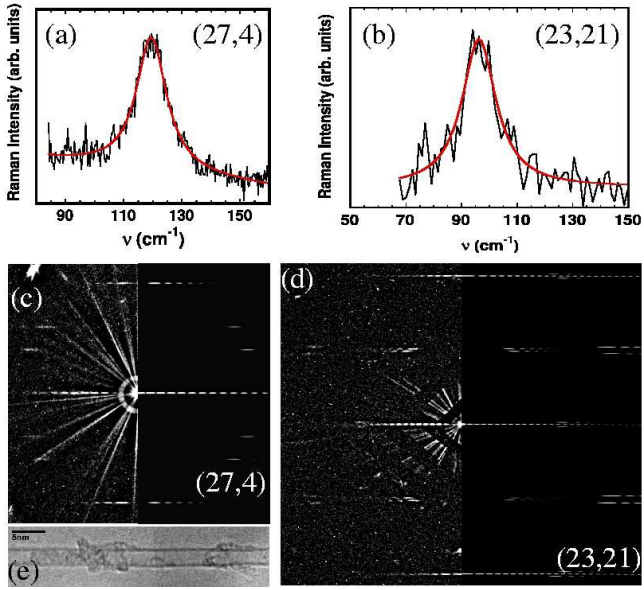}
\caption{(a) RBM of the (27,4) SWNT (E$_{Laser}$=1.92~eV), and (b) RBM of the (23,21) nanotube (E$_{Laser}$=1.6~eV). The electron diffraction patterns are shown in (c) and (d).
In each pattern the left half is the experimental image, while the right half is the simulated
one for comparison. (e) is a high-resolution image of the (27,4) SWNT.
\label{cap:3}}
\end{figure}

Figure \ref{cap:3} shows the Raman spectra and the diffraction patterns of the two tubes with the largest diameters of our data set, respectively the (27,4) and the (23,21) SWNTs.
The diameters of these semiconducting nanotubes are 2.287 and 2.984~nm (with a$_{C-C}$ = 0.142~nm) and their chiral angles are 6.8$^{\circ}$ and 28.5$^{\circ}$ respectively.
The RBM are located at 119~cm$^{-1}$ (FWHM = 11~cm$^{-1}$) and 95~cm$^{-1}$ (FWHM = 13~cm$^{-1}$) for the (27,4) and (23,21) tubes, respectively (Fig. \ref{cap:3}a and b).

\begin{figure}
\includegraphics[%
  width=1.0\linewidth]{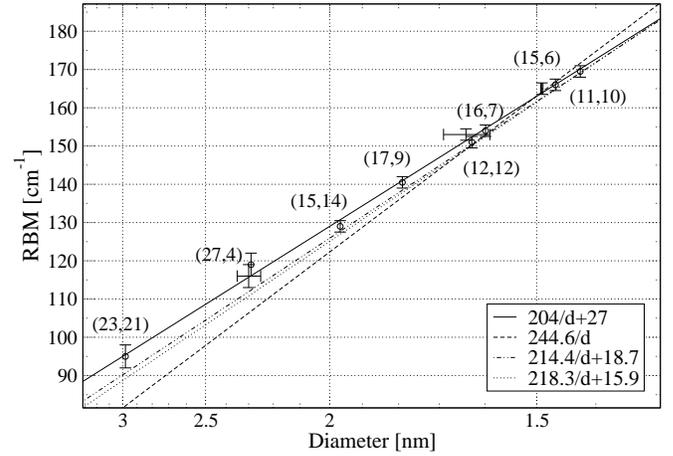}
\caption{RBM frequencies vs. nanotube diameter for nanotubes identified by
electron diffraction (see additional materials). Data points
marked with a circle are from unambiguously identified nanotubes, and the indices are given in the diagram. For the data points with a horizontal error bar, only a diameter estimate was possible.
Drawn as dashed lines are selected RBM vs. diameter relationships
from literature \cite{TelgPRL04,JorioPRL01,JorioPRB05}. The solid line best matches our data.\label{cap:4}}
\end{figure}

A huge number of experiments and modelization efforts were made to
relate the radial breathing mode (RBM) frequency to the nanotube structure.
A review and summary of various models and experiments is given in
\cite{sreich_CNTs-concepts+prop2004}. Our experimental approach provides
the first accurate and independent determination of the lattice structure of the tubes, (n,m) indices and then the diameters and chiral angles of all the tubes investigated, with their RBM frequencies. Figure \ref{cap:4} plots the experimental relation between the RBM frequencies and the nanotube diameters obtained by this way. We want to point out that these results are covering an unprecedented diameter range (1.4 nm to 3 nm).

We would like to precise again that the index assignment is solely
based on the electron diffraction analysis, and not on the Raman data.
The majority of the Raman spectra are recorded  on regions of micrometric size where only a single nanotube is detected by electron microscopy imaging.
In some cases, there are two nanotubes within or near of the laser spot.
We identify both of them and often find two very different diameters, so that only one of them is reasonable for the measured RBM frequency.
For example, the (17,9) SWNT (1.79~nm) is partly bundled with another tube with diameter >2.2~nm, which cannot exhibit the measured 140~cm$^{-1}$ RBM frequency.

On Figure \ref{cap:4}, all the measured points agree with a line obeying to the relation:  $\omega_{RBM}$=A/d+B, with A=204~cm$^{-1}$nm and B=27~cm$^{-1}$.
We point out that this RBM vs. diameter relation is measured on individual free-standing tubes, while in previously established relations the individual nanotubes were lying on a substrate \cite{JorioPRL01} or SDS wrapped \cite{TelgPRL04,FantiniPRL04b,JorioPRB05}.
Further, we have investigated a diameter range that was never studied before. Yet, we find a good agreement in the diameter range between 1.4 and 1.7~nm where our diameter range overlaps with the previous works. This suggests that
the previously established relationships can not be extrapolated to the large diameter nanotubes that are present in our samples. 


Our results allow to discuss the structural (n,m) determination of the Ref. \cite{JorioPRL01} only based on resonant Raman scattering data obtained by using the 1.58 eV laser excitation.
Indeed, we have measured the RBM frequencies of the (16,7) and (15,6) metallic tubes, and the RBM frequencies of these same tubes are given in Table I of the Ref. \cite{JorioPRL01}.
A comparison between the two data sets shows a complete agreement for the (16,7), since we found $\omega _{RBM}$ = 154 cm$^{-1}$ (E$_{laser}$ = 1.57 eV excitation).
Concerning the (15,6), the RBM is located at 166  cm$^{-1}$ (165 cm$^{-1}$ in  Ref. \cite{JorioPRL01}).
It can be pointed out that our spectrum is recorded by using the 1.7 eV laser excitation and at 1.58 eV in Ref. \cite{JorioPRL01}.
On this point our results seem supported by non-orthogonal tight-binding calculations
which predict a $\approx $0.1 eV relative difference for the separation energies
between van Hove singularities of these two nanotubes \cite{PopovPRB04}.
Once more, an exact determination of the resonant energies would however require a full set of laser lines \cite{TelgPRL04,FantiniPRL04b,JorioPRB05}.

The present combined electron diffraction and Raman scattering experiments on the same free-standing SWNT is a direct measurement of the relation between the RBM frequency and the tube diameter without modelizations of the vibrational
and electronic properties.
The main point of discussion concerns the dependence of the RBM frequency versus the tube diameter. For isolated tubes, the models predict a $d=\textrm{{A}}/\omega_{RBM}$ relation \cite{sreich_CNTs-concepts+prop2004}.
From the Raman experiments performed on SDS wrapped individual SWNTs \cite{TelgPRL04,FantiniPRL04b,JorioPRB05}, a $\omega_{RBM}=\textrm{A}/d+B$ relation was found. The non-zero value of the B parameter is commonly attributed to the effect of the environment.
Since most of the environmental influences are absent for the free-standing SWNTs investigated, this assumption seems questionable in our case.
Our results rather suggest that the dependence of RBM frequency with the inverse of the diameter might be slightly non-linear.
 Further, the precise agreement between RBM frequencies for identified tubes in SDS or on a substrate with our data on free-standing tubes shows that the influence of the environment is rather small.
To elucidate definitely this point, investigations on well characterized (n,m) free-standing nanotubes of small diameter are in progress.
The expected results will allow a direct comparison with more of the experimental results obtained from SDS wrapped individual SWNTs \cite{FantiniPRL04b,TelgPRL04,JorioPRB05} and  \textit{ab initio} calculations \cite{DubaiPRL02}.

In conclusion, we have obtained the Raman spectra of (n,m) nanotubes well characterized by electron diffraction and high-resolution TEM imaging.
We have directly measured the radial breathing mode frequency for a wide range of diameters.
Both the micro-Raman spectroscopy and the electron microscopic investigation are done on a freely suspended object without an influence from a substrate, surfactant or contacts.
Our measurements carried out on a wide range of diameters: from 1.4 nm up to 3 nm, confirm the previous established more or less model dependent laws in the range 1.4-1.7 nm in spite of the different environment. This raises questions on the interpretation of the environmental constant.
The unprecedented study of large diameter tubes shows that the RBM frequency is not simply inversely proportional to the nanotube diameter.


The measurement of vibrational modes for a precisely known structure can provide a direct test for molecular dynamics simulations. Further, we expect that the procedure shown here, due to the freely designable freestanding structure, can be adopted
to various nano-objects or macromolecules to combine electron microscopic structural analysis with Raman spectroscopy and potentially other investigations (transport, AFM) on the same object.

This work has been done in the framework of the GDRE n$^\circ $2756 ``Sciences and applications of the nanotubes - NANO-E''. The authors acknowledge financial support by the EU projects CARDECOM
and CANAPE and the BMBF project INKONAMI. We thank xlith.com for lithography
services. We thank P. Poncharal, A. Zahab, C. Koch, K. Hahn, M. Kelsch,
F. Phillipp and M. Rühle for support and helpful discussions.

\bibliographystyle{apsrev}
\bibliography{Raman,TEM,diverse,books,transport}

\end{document}